\documentclass[prd,a4paper,superscriptaddress,nofootinbib,10pt]{revtex4} 
\usepackage{tikz}
\usetikzlibrary{decorations.markings}
\usetikzlibrary{snakes}
\usepackage{graphicx}
\usepackage{amssymb}
\usepackage{amsmath}
\usepackage{lscape}
\usepackage{mathrsfs}
\usepackage{textcomp}
\usepackage{epsfig}
\usepackage{slashed}
\renewcommand{\d}{\mathrm{d}}

\begin{document}
\title{Noncommutative duality and fermionic quasinormal modes of the BTZ black hole}

\author{Kumar S. Gupta}
\email{kumars.gupta@saha.ac.in}
\affiliation{Theory Division, Saha Institute of Nuclear Physics, 1/AF Bidhannagar, Kolkata 700064, India}

\author{Tajron Juri\'c}
\email{tjuric@irb.hr}
\affiliation{Rudjer Bo\v{s}kovi\'c Institute, Bijeni\v cka  c.54, HR-10002 Zagreb, Croatia}
\affiliation{Instituto de Fisica, Universidade de Brasilia,
Caixa Postal 04455, 70919-970, Brasilia, DF, Brazil}

\author{Andjelo Samsarov}
\email{asamsarov@irb.hr}
\affiliation{Rudjer Bo\v{s}kovi\'c Institute, Bijeni\v cka  c.54, HR-10002 Zagreb, Croatia}

%\email{samsarov@unica.it}
%\affiliation{Dipartimento di Matematica e Informatica, Universit\`{a} di Cagliari, viale Merello 92, 09123 Cagliari, Italy 
% and INFN, Sezione di Cagliari}

\date{\today}

\begin{abstract}

We analyze the fermionic quasinormal modes of the BTZ black hole in the presence of space-time noncommutativity. Our analysis exploits a duality between a spinless and spinning BTZ black hole, the spin being proportional to the noncommutative deformation parameter. Using the AdS/CFT correspondence we show that the horizon temperatures in the dual CFT are modified due to noncommutative contributions. We demonstrate the equivalence between the quasinormal and non-quasinormal modes for the noncommutative fermionic probes, which provides further evidence of holography in the noncommutative setting. Finally we present an analysis of the emission of Dirac fermions and the corresponding tunneling amplitude within this noncommutative framework.

\end{abstract}

\maketitle

\section{Introduction}

Quasinormal modes (QNM) of black holes \cite{rg,vish,press,chandra,Cardoso:2001hn,BTZ,cardosoreview,rew} provide key signatures of the gravitational waves. The QNM's arising from the perturbation of a black hole depend only on the parameters of the black hole and not on the details of the perturbation. It is this feature that makes the QNM's a fundamental quantity in exploring properties of black holes. The recent experimental discovery of gravitational waves including the ringdown phase  arising from black hole mergers \cite{ligo} have opened up new possibilities for the observations of the QNM's and it provides an impetus to the idea that gravitational wave astronomy may in the future be used to probe of the primordial universe at the Planck scale.

It is widely believed that properties of the space-time at the Planck scale could be very different from what we observe today. There are various models of such space-times including string theory \cite{Aharony}, loop gravity \cite{carlo} and noncommutative  geometry \cite{cones}, all of which suggest that the space-time might have some discrete structure at the quantum gravity scale. In particular, it is known that general relativity and the quantum uncertainty principle together predict a very general class of noncommutative space-times \cite{ahluwalia,dop1,dop2}. Furthermore it has been shown that the space-times associated with a  variety of black holes at the Planck scale could be described by a $\kappa$-Minkowski algebra \cite{btzkappa,ohl}. In this paper we shall take the $\kappa$-Minkowski algebra \cite{kappa1, kappa2, kappa3, KowalskiGlikman:2002we, KowalskiGlikman:2002jr, Dimitrijevic:2003wv, Meljanac:2006ui, KresicJuric:2007nh, Borowiec:2008uj, Meljanac:2007xb} as the prototype of a space-time at the Planck scale and shall investigate various features of QNM's and associated physics in that background.

In a previous set of works \cite{1,2}, the properties of a noncommutative (NC) $\kappa$-Minkowski scalar field in the background of a BTZ black hole \cite{banados} were investigated and elaborated further in \cite{ent}. It was shown that probing a spinless BTZ black hole with a $\kappa$-Minkowski scalar field is equivalent to probing a spinning BTZ black hole with a commutative scalar field \cite{2}. This result was established by showing that the Klein-Gordon equations in these two situations are identical upto the first order in the NC deformation parameter. The effective spin of the dual BTZ black hole was obtained from the corresponding black hole entropy \cite{2}, which depends on the NC parameter and captures the back reaction of the NC scalar field on the BTZ space-time.  The restriction of the analysis only up to the first order in the deformation parameter is prompted by two main considerations. First, the noncommutativity is a Planck scale effect, which in the present epoch would be very small. Hence from a phenomenological point of view, the first order effects would be most dominant. Furthermore, the full noncommutative equations of motion are extremely complicated \cite{1}, and only by restricting the analysis up to the first order we could obtain analytical results. In this paper we explore further consequences of this duality and analyze the fermionic QNM's in this dual BTZ space-time. 

In addition to traditional derivation of Hawking radiation \cite{hawkingmain}, there exists an alternative approach towards understanding  black hole radiation and the physical processes that lie behind. This approach is based on a semi-classical method of modeling Hawking radiation as a tunneling effect from the inside to the outside of the horizon \cite{kraus-wilczek,parikh-wilczek}. The procedure amounts to calculating the imaginary part of the classical action which can be shown to fix the tunneling probability amplitude. On the other hand the classical action itself can be calculated either by
null-geodesic method \cite{parikh-wilczek} or by Hamilton-Jacobi method \cite {angheben,padmanabhan}. In this paper we apply the tunneling framework in order to investigate the impact that NC nature of space-time  might have on  the tunneling probability for the classically forbidden trajectory of fermions passing from the inside to the outside of the horizon.

The paper is structured as follows. In Section II, the NC duality is presented. It is based on the observation that the equation of motion for a NC scalar field  in the background of a spinless BTZ black hole can be rewritten in a form of a KG-equation for a commutative scalar field, but now moving in the background of a dual  BTZ black hole with non-zero spin or angular momentum. Section III discusses a derivation of  Dirac equation in NC setting by taking the ``square-root'' of  NC KG-equation through the use of NC duality. Furthermore, new contributions to the fermionic QNM's are found from the NC effects. In Section IV we discuss the  relevance of holography and QNM's within the NC framework. In Section V, we calculate the probability amplitude for quantum tunneling using the WKB which allows us to obtain the corresponding Hawking temperature. We conclude the paper in Section VI with some comments.

%%%%%%%%%%%%%%%%%%%%%%%%%%%%%%%%%%%%%%%%%%%%%%%%%%%%%%%%%%%%%%%%%%%%%%%%%
%%%%%%%%%%%%%%%%%%%% %%%%%%%%%%%%%%%%%%%%%
%%%%%%%%%%%%%%%%%%%%%%%%%%%%%%%%%%%%%%%%%%%%%%%%%%%%%%%%%%%%%%%

\section{NC duality}

As mentioned in the Introduction, a spinless BTZ black hole being probed with a $\kappa$-NC scalar field is dual to a spinning BTZ black hole probed with a commutative scalar field \cite{2}. This equivalence together with the expression of the corresponding BTZ black hole entropy allows us to identify the spin of the dual BTZ black hole which depends on the NC parameter. We briefly review how to derive the spin of the dual black hole which will be used later in this paper.

 A massless NC scalar particle in the background 
\begin{equation}\label{btzmetric}
g'_{\mu\nu}=\begin{pmatrix}
M - \frac{r^2}{l^2} &0&0\\
0&\frac{1}{\frac{r^2}{l^2}-M}&0\\
0&0&r^2\\
\end{pmatrix},
\end{equation}
of the BTZ black hole with mass $M$ and angular momentum $J=0$ is described by the equation which can be symbolically presented as
\begin{equation} \label{1}
  ({\Box_{g'}} + {\mathcal{O}}(a)) \Phi = 0.
\end{equation}
The parameter $l$ is related to the cosmological constant $\Lambda$ as $l = \sqrt{-\frac{1}{\Lambda}}$ and $a$ is the deformation parameter, $a=\frac{1}{\kappa},$ that sets up the NC scale, commonly related to the Planck length. ${\Box_{g'}}$ is the KG operator in the metric (\ref{btzmetric}). The second term in the above equation is a generic expression  representing a whole set  of corrections  induced by the noncommutative nature of spacetime. It was shown in \cite{2} that equation (\ref{1}) may be rewritten in the form 
\begin{equation} \label{7}
  ({\Box_{g}} - m^2) \Phi = 0,
\end{equation}
where $\Box_{g}$ is the
Klein-Gordon operator for the metric
\begin{equation}\label{eqbtzmetric}
g_{\mu\nu}=\begin{pmatrix}
 M^d - \frac{r^2}{l^2} &0& \frac{-J^d}{2}\\
 0&\frac{1}{\frac{r^2}{l^2} + \frac{{(J^{d})}^2}{4 r^2} -M^d}&0\\
\frac{-J^d}{2} &0& r^2\\
\end{pmatrix},
\end{equation}
describing a geometric background of the BTZ black hole with mass $M^d$ and angular momentum $J=J^d.$ Moreover, in the commutative dual picture the scalar particle has acquired the mass $m$ (see appendix A and \cite{ent} for more details). 
 Different black hole parameters in the dual picture indicate that the black hole in the new setting  was set into a rotational motion with  angular  momentum $J = J_d$. This appears to be possible  due to the effects of noncommutative geometry which  enable a noncommutative probe to influence the geometry through which it propagates, thus making the instance for the backreaction mechanism in  this particular situation.

 It has to be noted that the term backreaction here appears in the same sense as for example in \cite{Martinez:1996uv} where it  represents a situation where the propagating matter modifies the geometry that  it probes. However, there is also an important difference. While in \cite{Martinez:1996uv}, a material content of the propagating particles itself is a source of the geometry modification,   in our case  the reason for the change in the geometry is contained within a noncommutative nature of spacetime at the Planck scale. In this way,  NC nature of spacetime, in particular its grainlike structure, acts as an agent  that mediates the influence of the propagating matter toward a black hole background, without modifying the  energy-momentum tensor.

In order to obtain the specific expression for $J^d$, consider the entropy of the spinless BTZ black hole as probed by the NC scalar field \cite{1}, which is given by
\begin{equation} \label{snc}
S^{NC}=\frac{A_0}{4G}\left(1+a\beta\sqrt{M}\frac{8\pi\zeta(2)}{3l\zeta(3)}\right),
\end{equation}
where $A_0=2\pi l\sqrt{M}$ is the area of the spinless BTZ with mass $M$. The entropy of a black hole can be obtained from the solutions of the corresponding Klein-Gordon equation \cite{gthooft,1}. The main steps leading to the entropy are also described in appendix A. Since the Klein-Gordon equations for the spinless BTZ and its spinning dual are identical, we can postulate the equivalence $S^{NC}=S^d$, where
\begin{equation}
S^d=\frac{A^d}{4G}, \quad A^d=2\pi r_+ , 
\end{equation}
where $S^d$ is the entropy of the dual BTZ black hole and 
\begin{equation} \label{rplus} 
 r_{+}=\frac{l\sqrt{M}}{\sqrt{2}}\sqrt{1+\sqrt{1-\frac{(J^d)^2}{M^2l^2}}},
\end{equation}
is the outer horizon of the dual BTZ black hole. These conditions imply that
\begin{equation}\label{Aovi}
  A^d = 2\pi l\sqrt{M}\left(1+a\beta\sqrt{M}\frac{8\pi\zeta(2)}{3l\zeta(3)}\right),
\end{equation}
which gives
\begin{equation}\label{J}
(J^d(a))^2=\lambda\frac{64}{3}\pi\frac{\zeta(2)}{\zeta(3)}lM^{5/2}+O(a^2),
\end{equation}
where the abbreviation  $\lambda=-a\beta$ has been used. Since $a\beta\in\mathbb{R} \backslash \left\{0\right\}$, we  restrict realizations\footnote{For details see \cite{1,2}.} to be of those types  where $a\beta < 0$. We shall use this expression for the $J^d$ in the subsequent analysis.

We conclude this Section with the following observations. One could in principle find other NC dualities where the dual BTZ would also acquire a change in its mass. However, it is easy to see that in this case the corrections in the mass would be of higher order then those in spin. Namely, if we look at the extreme  dual picture where $J^d=0$ and the whole NC effect is squeezed into $M^d,$ we find (see appendix B)
\begin{equation} \label{md}
M^d=M\left(1+a\beta \frac{\sqrt{M}}{l} \frac{16\pi}{3}\frac{\zeta(2)}{\zeta(3)}\right)+O(a^2),
\end{equation}
which will lead to  higher order corrections in the physically observable quantities such as the QNM frequencies and the greybody factor. Since the mass $M^d$ in (\ref{md}) pertains to a particular physical situation, which deserves a special analysis (see  appendix B), we ascribe to it a special designation,  $M^f$.

We can find a bound on NC parameter $\lambda(a)$ from the condition
\begin{equation}
1-\frac{(J^d)^2}{M^2 l^2}>0,
\end{equation}
which leads to
\begin{equation}\label{uvjet}
\lambda(a) < {l \over {c \sqrt{M}}},~~~~~c = \frac{64 \pi}{3} \frac{\xi(2)}{\xi(3)}
\end{equation}
 For any macroscopic black hole, eqn. (\ref{uvjet}) easily fulfills the condition that $\lambda(a)\approx\frac{1}{M_{Planck}}$. Moreover, it allows for the limit $a \rightarrow 0$, when we recover the commutative results.

\section{Fermionic QNM}

%\subsection{Dirac QNM}
One way to deal with  Dirac particles in  NC setting might be to pursue an analysis similar to that 
carried out in the Appendix A and ref.\cite{1}, though appropriately accommodated to fermions. This is however a demanding and still ambiguous procedure since it involves a star-product between fermions which is yet not fully developed, even in flat spacetime. Therefore, we propose to skip this rather involved procedure and focus on the approach based on NC duality. By way of, since we were able to show that the NC picture of KG particle on BTZ $(M,J=0)$ is equivalent to a massive KG particle on a dual commutative BTZ with angular momentum $\left|J^d\right|\propto\sqrt{a\beta}$, we can search for the ``square-root'' $~D~$ of the corresponding KG operator $~\Box_g~$ in the sense that the relation $~D^2 = \Box_g~$ holds. Proceeding this way, we are able to claim that the operator $~D~$ thus found defines  a  Dirac equation that governs a fictitious commutative picture, which in a way analogous to the bosonic case, pertains to a noncommutative Dirac particle  on the BTZ background. 
Although we are not able to say what actual situation does it describe (whether it be NC fermion field coupled to commutative background or commutative fermion field coupled to NC background or both)
 the equation so obtained 
% is exactly what we  have searched for.
  in a sense gives an effective description of  Dirac fermions on the BTZ background in  NC setting, realized in terms of the usual commutative narrative. Therefore, the whole problem  reduces  to finding a Dirac equation on the general BTZ background $(M,J\propto\sqrt{a\beta}), $ which in the standard coordinates
$~ x^{\mu} = (t,r,\phi)~$ is given by
\begin{equation} \label{btzoriginal}
  \d s^2 = - \bigg(  \frac{r^2}{l^2} - M + \frac{J^2}{4r^2} \bigg) \d t^2 + \frac{\d r^2}{\frac{r^2}{l^2} - M 
    + \frac{J^2}    {4r^2}}  + r^2  {\bigg(  \d  \phi - \frac{J}{2r^2} \d t  \bigg)}^2.
\end{equation}
Dirac equation in BTZ background has also been considered in \cite{Dasgupta:1998jg,Das:1999pt,Kerner,3,5,6,7}.
% This is already done (see eqs.(1-37) in \cite{3}, or eqs.(3-8) in \cite{4} or \cite{5,6}).

It however appears more convenient to set up the equation of motion in the basis of coordinates
$~ x^{\mu} = (x^+, \rho, x^-),~$ that are linked with the standard coordinates $\; t, r, \phi  \;$ through the transformation relations
\begin{equation} \label{transformations}
  x^+ = \frac{1}{l}r_+ t - r_- \phi,  \quad  \quad x^- = r_+ \phi -  \frac{1}{l} r_-   t   \quad  \quad   \mbox{and} \quad \quad  \tanh \rho = \sqrt{\frac{r^2 - r_+^2}{r^2 - r_-^2}}.
\end{equation}
In these coordinates the metric (\ref{btzoriginal}) looks as
\begin{equation}
\d s^2=-\sinh^2 \rho ~ {(\d x^+)}^2 + l^2 \d \rho^2 + \cosh^2 \rho ~ {(  \d x^- )}^2.
\end{equation}
This gives
\begin{equation} \label{metric}
  g_{\mu \nu} =
\left( \begin{array}{ccccc}
  -\sinh^2 \rho & 0  & 0  \\
   0   & l^2 & 0  \\ 
   0  & 0 & \cosh^2 \rho &  \\
\end{array} \right)  \quad \quad \mbox{and for the inverse metric}  \quad \quad
 g^{\mu \nu} =
\left( \begin{array}{ccccc}
  -\frac{1}{\sinh^2 \rho} & 0  & 0  \\
   0   & \frac{1}{l^2} & 0  \\ 
   0  & 0 & \frac{1}{\cosh^2 \rho } &  \\
\end{array} \right)
\end{equation}

By linearizing (or taking the Dirac ``square-root'') of the KG-equation  (\ref{7}) in a way already described,  we get the Dirac equation in the BTZ background
\begin{equation}\label{diracnovo}
(  \gamma^a \nabla_a + m)\Psi=0,
\end{equation}
where the Latin indices such as $a, ~(a =0,1,2)$ refer to intrinsic coordinates and $\gamma^a$ are the standard flat space Dirac gamma matrices, $~\{ \gamma_a, \gamma_b \} = 2 \eta_{ab},~$ where
\begin{equation} \label{metriceta}
  \eta_{ab} = \eta^{ab} =
\left( \begin{array}{ccccc}
  -1  & 0  & 0  \\
   0   & 1 & 0  \\ 
   0  & 0 & 1 &  \\
\end{array} \right).  
\end{equation}

The Dirac operator  $~  \gamma^a \nabla_a ~$ on a curved space
 is introduced in terms of tetrads (dreibeins) $~ e^a_{~~\mu} ~$ and their inverse  $~ e_a^{~~\mu}, ~$ satisfying
$~ e^a_{~~\mu}  e_a^{~~\nu} = \delta_{\mu}^{~~\nu} ~$ and  $~ e^a_{~~\mu}  e_b^{~~\mu} = \delta^{a}_{~~b}. ~$ Moreover, they also satisfy 
$~g_{\mu \nu} =  e^a_{~~\mu}  e^b_{~~\nu}  \eta_{ab} ~$ and $~g^{\mu \nu} =  e_a^{~~\mu}  e_b^{~~\nu}  \eta^{ab}.$  
In what follows we use the setting defined in \cite{Dasgupta:1998jg}.
This setting consists of  the dreibein frame  chosen to be
\begin{equation} \label{metrictetrad}
  e^a_{~~\mu} =
\left( \begin{array}{ccccc}
  \sinh \rho & 0  & 0  \\
   0   & l & 0  \\ 
   0  & 0 & \cosh \rho &  \\
\end{array} \right)  \quad \quad \mbox{with the corresponding  inverse matrix}  \quad \quad
 e_a^{~~\mu} =
\left( \begin{array}{ccccc}
  \frac{1}{\sinh \rho} & 0  & 0  \\
   0   & \frac{1}{l} & 0  \\ 
   0  & 0 & \frac{1}{\cosh \rho } &  \\
\end{array} \right)
\end{equation}
and  the following representation of gamma matrices
\begin{equation} \label{gammarep}
  \gamma^0 = i \sigma_2 =
\left( \begin{array}{ccccc}
  0  & 1  \\
   -1   & 0  \\ 
\end{array} \right),  \quad \quad 
 \gamma^1 =  \sigma_1 =
\left( \begin{array}{ccccc}
  0  & 1  \\
   1   & 0  \\  
 \end{array} \right), \quad \quad 
 \gamma^2 =  \sigma_3 =
\left( \begin{array}{ccccc}
  1  & 0  \\
   0   & -1  \\ 
\end{array} \right),  
\end{equation} 
where $~ \sigma_i, ~ (i=1,2,3)~$ are the usual Pauli matrices.

By writing out a detailed structure of the covariant derivative $~\nabla_a, ~$  the Dirac equation (\ref{diracnovo}) takes the form
\begin{equation}\label{diracnovo1}
\bigg[  \gamma^a e_a^{~~\mu} \bigg( \partial_{\mu} - \frac{i}{2} {\omega}_{\mu}^{~~cd} \Sigma_{cd} \bigg) + m  \bigg]\Psi=0.
\end{equation}
Here $~\Sigma_{cd} = \frac{i}{4} [\gamma_c, \gamma_d]~$
and the coefficients of the spin connection $~\omega_{\mu}^{~~ab}~$ are given by
\begin{equation} \label{spinconnection}
\begin{split} 
  \omega_{\mu}^{~~ab} & = e^a_{~~\nu} \eta^{bc} \partial_{\mu} e_c^{~~\nu} + e^a_{~~\nu} \eta^{bc} e_c^{~~\lambda} \Gamma^{\nu}_{~~\mu \lambda} \\
  & = \frac{1}{2} e^{a\nu} \bigg( \partial_{\mu} e^b_{~~\nu} - \partial_{\nu} e^b_{~~\mu} \bigg)
 - \frac{1}{2} e^{b\nu} \bigg( \partial_{\mu} e^a_{~~\nu} - \partial_{\nu} e^a_{~~\mu} \bigg)
 - \frac{1}{2} e^{a\rho} e^{b\sigma} \bigg( \partial_{\rho} e_{c\sigma} - \partial_{\sigma} e_{c\rho} \bigg) e^c_{~~\mu},   \nonumber
\end{split}
\end{equation}
where $~\Gamma^{\nu}_{~~\mu \lambda} = \frac{1}{2} g^{\nu \delta} \bigg( \partial_{\mu} g_{\delta \lambda} + 
  \partial_{\lambda} g_{\mu \delta} - \partial_{\delta} g_{\mu \lambda} \bigg)~$ are the coefficients of the affine connection.

Straightforward calculation shows that the only coefficients of affine connection that are different from zero are
\begin{equation}
\begin{split}
 & \Gamma^{2}_{~~12} = \Gamma^{2}_{~~21} \equiv \Gamma^{x^-}_{~~ x^- \rho} = \tanh \rho, \quad \Gamma^{0}_{~~01} = \Gamma^{0}_{~~10} \equiv \Gamma^{x^+}_{~~ x^+ \rho} = \frac{1}{\tanh \rho}, \\
  & \Gamma^{1}_{~~00}  \equiv \Gamma^{\rho}_{~~ x^+  x^+} = \frac{\sinh (2\rho)}{2 l^2}, \quad \Gamma^{1}_{~~22}  \equiv \Gamma^{\rho}_{~~ x^-  x^-} = -\frac{\sinh (2\rho)}{2 l^2}, 
\end{split}
\end{equation}
With $~e^0_{~~x_+} = \sinh \rho, ~e^1_{~~\rho} = l, ~ e^2_{~~x_-} = \cosh \rho,$ and others being $0,$ one can calculate the only non zero components of the spin connection
\begin{equation} \label{spinconnection1}
 \begin{split}
 & \omega_{0}^{~~01}    \equiv \omega_{x^+}^{~~ 01} =
   -\omega_{0}^{~~10} =
 \frac{1}{l} \cosh \rho, \\
  & \omega_{2}^{~~12}    \equiv \omega_{x^-}^{~~ 12} =
   -\omega_{2}^{~~21} =
 -\frac{1}{l} \sinh \rho, 
\end{split}
\end{equation}
leading to the Dirac equation
\begin{equation} \label{diracnovo2}
 \bigg[  \frac{1}{l} \gamma^1 \bigg( \frac{\partial}{\partial \rho}  + \frac{\cosh \rho}{2\sinh \rho} + \frac{\sinh \rho}{2\cosh \rho} \bigg)  + \gamma^0 \frac{1}{\sinh \rho}   
   \frac{\partial}{\partial x^+} 
 +  \gamma^2 \frac{1}{\cosh \rho} \frac{\partial}{\partial x^-} + m \bigg] \Psi=0.
\end{equation}

 In order to solve this equation for the wavefunction $~\Psi \equiv \Psi(r,t,\phi)=\Psi(\rho, x^+, x^-), ~$ we  follow \cite{Dasgupta:1998jg},\cite{7} and proceed with a series of steps which begins by taking the ansatz
\begin{equation} \label{anzatz}
 \begin{split}
\Psi &=
  \left(\begin{matrix} \psi_1  \\ \psi_2  \\   \end{matrix}\right)
  =\left(\begin{matrix} \psi_1 (r) \\ \psi_2 (r) \\ \end{matrix}\right)\text{exp}\left[-\frac{i}{l}(\omega t - j \phi)\right]
= \left(\begin{matrix} \psi_1 (\rho) \\ \psi_2 (\rho) \\ \end{matrix}\right)\text{exp}\left[-i(k_+ x^+ +k_- x^-)\right] \\
 &= \frac{1}{2} \left(\begin{matrix} P (\rho) + Q(\rho) \\  P (\rho) - Q(\rho) \\ \end{matrix}\right)\text{exp}\left[-i(k_+ x^+ +k_- x^-)\right],
\end{split}
\end{equation}
where $\omega$ and $j$ are respectively the energy and angular momentum of the spin $1/2$ particle and they are related to $k_+$ and $k_-$ as
\begin{equation}  \label{kpluskminus}
  k_+ = \frac{l\omega  r_+ - j r_-}{l(r_+^2 - r_-^2)}, \quad
  k_- = \frac{l\omega r_- - j r_+}{l(r_+^2 - r_-^2)}.
\end{equation}
Inserting  (\ref{anzatz}) into (\ref{diracnovo2}) 
%%%%%%%%%%%%%%%%%%%% leads to
% 	consistent with   gamma^a nabla_a + m
%\begin{equation} \label{pairofeqs}
% \begin{split}
%   & \bigg( \frac{\partial}{\partial \rho}  + \frac{\cosh \rho}%{2\sinh \rho} + \frac{\sinh \rho}{2\cosh \rho} \bigg) P(\rho)
%   + \frac{ilk_+}{\sinh \rho} Q(\rho) - \frac{ilk_-}{\cosh \rho} Q(\rho) + lm P(\rho) =0,
% \\
% & \bigg( \frac{\partial}{\partial \rho}  + \frac{\cosh \rho}{2\sinh \rho} + \frac{\sinh \rho}{2\cosh \rho} \bigg) Q(\rho)
%   + \frac{ilk_+}{\sinh \rho} P(\rho) + \frac{ilk_-}{\cosh \rho}  P(\rho) - lm Q(\rho) =0.
%\end{split}
%\end{equation}
% 	consistent with  i gamma^a nabla_a - m
%   m --> im
%\begin{equation} \label{pairofeqss}
% \begin{split}
%   & \bigg( \frac{\partial}{\partial \rho}  + \frac{\cosh \rho}{2\sinh \rho} + \frac{\sinh \rho}{2\cosh \rho} \bigg) P(\rho)
%   + \frac{ilk_+}{\sinh \rho} Q(\rho) - \frac{ilk_-}{\cosh \rho} Q(\rho) + ilm P(\rho) =0,
% \\
% & \bigg( \frac{\partial}{\partial \rho}  + \frac{\cosh \rho}{2\sinh \rho} + \frac{\sinh \rho}{2\cosh \rho} \bigg) Q(\rho)
%   + \frac{ilk_+}{\sinh \rho} P(\rho) + \frac{ilk_-}{\cosh \rho}  P(\rho) - ilm Q(\rho) =0.
%\end{split}
%\end{equation}
%Then, after using the substitutions
%%%%%%%%%%%%%%%%%%%
and using afterwords the following sequence of substitutions 
\begin{equation} \label{substitution}
  P(\rho) =\sqrt{\frac{\cosh \rho + \sinh \rho}{\cosh \rho ~\sinh \rho }} ~P'(\rho), 
  \quad  
  Q(\rho) = \sqrt{\frac{\cosh \rho - \sinh \rho}{\cosh \rho ~\sinh \rho }} ~Q'(\rho), \quad  z=\tanh^2 \rho,
\end{equation}
the  set of equations contained in (\ref{diracnovo2}) reduces to 
% 	consistent with   gamma^a nabla_a + m
%\begin{equation} \label{pairofeqs2}
%\begin{split}
%   & 2 \sqrt{z} (1-z) \frac{d}{dz} \tilde{P'} (z) + il \bigg(         
%  k_+ \frac{1}{\sqrt{z}} + k_- \sqrt{z} - (k_+ + k_-) \bigg) %\tilde{Q'} (z) + \bigg(\frac{1}{2}+ lm \bigg) \tilde{P'} (z) =0,
% \\
% & 2 \sqrt{z} (1-z) \frac{d}{dz} \tilde{Q'} (z) + il \bigg(         
%  k_+ \frac{1}{\sqrt{z}} + k_- \sqrt{z} + (k_+ + k_-) \bigg) %\tilde{P'} (z) - \bigg(\frac{1}{2}+ lm \bigg) \tilde{Q'} (z) =0.
%\end{split}
%\end{equation}
% 	consistent with  i gamma^a nabla_a - m
%   m --> im
%\begin{equation} \label{pairofeqs2}
%\begin{split}
%   & 2 \sqrt{z} (1-z) \frac{d}{dz} \tilde{P'} (z) + il \bigg(         
%  k_+ \frac{1}{\sqrt{z}} + k_- \sqrt{z} - (k_+ + k_-) \bigg) \tilde{Q'} (z) + \bigg(\frac{1}{2}+ ilm \bigg) \tilde{P'} (z) =0,
% \\
% & 2 \sqrt{z} (1-z) \frac{d}{dz} \tilde{Q'} (z) + il \bigg(         
%  k_+ \frac{1}{\sqrt{z}} + k_- \sqrt{z} + (k_+ + k_-) \bigg) \tilde{P'} (z) - \bigg(\frac{1}{2}+ ilm \bigg) \tilde{Q'} (z) =0.
%\end{split}
%\end{equation}
\begin{equation} \label{pairofeqs2}
 \begin{split}
   & 2 \sqrt{z} (1-z) \frac{d}{dz} P' (z) + il \bigg(         
  k_+ \frac{1}{\sqrt{z}} + k_- \sqrt{z} - (k_+ + k_-) \bigg) Q' (z) + \bigg(\frac{1}{2}+ lm \bigg) P' (z) =0,
 \\
 & 2 \sqrt{z} (1-z) \frac{d}{dz} Q' (z) + il \bigg(         
  k_+ \frac{1}{\sqrt{z}} + k_- \sqrt{z} + (k_+ + k_-) \bigg) P' (z) - \bigg(\frac{1}{2}+ lm \bigg) Q' (z) =0.
\end{split}
\end{equation}
By further putting $~ P' = \psi'_1 + \psi'_2~$ and $~ Q' = \psi'_1 - \psi'_2~$ one arrives at
\begin{equation}\label{diracap1}
 \begin{split} 
 & 2 \sqrt{z}(1-z) \frac{d}{dz} \psi^{\prime}_1 + il\bigg( k_+  \frac{1}{\sqrt{z}} +k_- \sqrt{z} \bigg) \psi^{\prime}_1 + \left[il(k_+ +k_-)+ lm +\frac{1}{2}\right]\psi^{\prime}_{2} =0, \\
 & 2 \sqrt{z}(1-z) \frac{d}{dz} \psi^{\prime}_2 - il\bigg( k_+  \frac{1}{\sqrt{z}} +k_- \sqrt{z} \bigg) \psi^{\prime}_2 - \left[il(k_+ +k_-)- lm -\frac{1}{2}\right]\psi^{\prime}_{1} =0.
\end{split}
\end{equation}

Combining two last  equations gives rise to the second order differential equation
\begin{equation}  \label{2nddiffeq}
  z(1-z) \frac{d^2}{dz^2} \psi'_1 + \frac{1-3z}{2} \frac{d}{dz} \psi'_1 + \frac{1}{4} \bigg[  \frac{l^2 k_+^2 - il k_+}{z} + ilk_- -l^2 k_-^2 - \frac{{(lm + \frac{1}{2})}^2}{1-z}\bigg] \psi'_1  =0.
\end{equation}
The solution to this equation is
\begin{equation}  \label{psi1sol}
 \psi^{\prime}_{1}=z^{\alpha}(1-z)^{\beta} \bigg( C_1 F(a, b, c; z)  + C_2 z^{1-c} F(a-c+1, b-c+1, 2-c; z) \bigg),
\end{equation}
where $C_1$ and $C_2$ are the integration constants and
\begin{equation}\begin{split}
&  \alpha=-\frac{ilk_+}{2}, \quad \beta =-\frac{1}{2}\left(lm +\frac{1}{2}\right), \quad c = 2\alpha +\frac{1}{2},              \\
&  a = \alpha + \beta + \frac{ilk_-}{2} + \frac{1}{2} = \frac{l (k_+ -k_-)}{2i}+\beta  + \frac{1}{2}, \\
&  b =  \alpha + \beta - \frac{ilk_-}{2} = \frac{l (k_+ +k_- )}{2i}+\beta.
\end{split}\end{equation}
While the first linearly independent solution in (\ref{psi1sol})  behaves as $~ z^{\alpha} = e^{-\frac{i}{2}lk_+ \ln z },~$ the second one behaves as $~  z^{\alpha + 1-c} \sim  e^{\frac{i}{2}lk_+ \ln z }.~$ Moreover, since we are interested in quasinormal modes, which are in turn  characterized by a purely in-going flux at the horizon, not the whole solution in (\ref{psi1sol})  is of interest to us. This means that the latter boundary condition, requiring  the wave function to be purely in-going at the horizon, further 
fixes the form of $\psi'_1$ in (\ref{psi1sol}).
On top of that, the same procedure can be performed for the second component $\psi'_2$ of the spinor wave function, which finally yields the following set of solutions for the components $\psi'_1$ and $\psi'_2,$
\begin{equation}\begin{split}  \label{psi1sol1}
&\psi^{\prime}_{1}=z^{\alpha}(1-z)^{\beta}F(a, b, c; z),  \\
&\psi^{\prime}_{2}=\left(\frac{a -c}{c}\right)z^{\alpha+\frac{1}{2}}(1-z)^{\beta}F(a, b +1, c  +1; z),
\end{split}\end{equation}
having a purely in-going flux at the horizon.

 Another boundary condition specifying quasinormal modes is that of vanishing flux at infinity, $\rho \rightarrow \infty$
(or equivalently, $z \rightarrow 1$).  The flux ${\mathcal{J}}_{\rho}$ in the radial $\rho$ direction is given by
\begin{equation}  \label{flux}
{\mathcal{J}}_{\rho}=\sqrt{-g}\bar{\Psi}\gamma_{\rho}\Psi
  = \sqrt{-g}\bar{\Psi} e^{a}_{~~\rho}  \gamma_{a}\Psi
\end{equation}
and  by using Eqs.(\ref{metric}),(\ref{metrictetrad}),(\ref{anzatz}) and (\ref{substitution})  it can be explicitly  calculated  as
\begin{equation}  \begin{split} \label{flux1}
& {\mathcal{J}}_{\rho}=  \frac{l^2 \sqrt{z}}{1-z}  \bigg[ \frac{1}{2} \frac{1+ \sqrt{z}}{\sqrt{z}} {(1-z)}^{1/2} {P'}^{*} (z) P'(z) - 
      \frac{1}{2} \frac{1- \sqrt{z}}{\sqrt{z}} {(1-z)}^{1/2} {Q'}^{*} (z) Q'(z) \bigg] \\
&  =  \frac{l^2 \sqrt{z}}{1-z}  \frac{{(1-z)}^{1/2}}{\sqrt{z}} \bigg[ {\psi'_1}^{*} \psi'_2 + {\psi'_2}^{*} \psi'_1 + \sqrt{z} \bigg(  {\psi'_1}^{*} \psi'_1 + {\psi'_2}^{*} \psi'_2  \bigg)  \bigg]
  =  \frac{l^2 }{ \sqrt{1-z}} \bigg[ {\psi'_1}^{*} \psi'_2 + {\psi'_2}^{*} \psi'_1 + \sqrt{z} \bigg(  {\psi'_1}^{*} \psi'_1 + {\psi'_2}^{*} \psi'_2  \bigg)  \bigg],
\end{split}
\end{equation}
where $\psi'_1$ and $\psi'_2$ are given by (\ref{psi1sol1}).

According to (\ref{psi1sol1}) and due to $\beta $ being real, each of the terms in the square brackets produces the additional  factor of ${(1-z)}^{2\beta},$ which together with the prefactor of ${(1-z)}^{-1/2}, $ standing in front of the square brackets in (\ref{flux1}), leads to a singular behavior of the flux in the far asymptotic region. In order to fully characterize the flux in the region $z \rightarrow 1,$ we need to expand the hypergeometric functions that make up solutions in (\ref{psi1sol1}) around this point. As it will be seen shortly, this expansion will no further aggravate the  divergence structure of the flux far from the horizon.

 The solutions (\ref{psi1sol1}) are the exact solutions which are valid in all regions outside of the outer horizon. In exactly the same way  the expression in (\ref{flux1})  represents an exact flux of fermionic particles which is  valid throughout the space. Yet, since we are interested in the behavior of the flux at  far infinity, we use the linear transformation formula
\begin{eqnarray}  \label{lintransf}
  F(a,b,c;z)  & = & \frac{\Gamma(c) \Gamma(c-a-b)}{\Gamma(c-a) \Gamma(c-b)} F(a,b,a+b-c+1; 1-z)  \nonumber \\
     &  +  & {(1-z)}^{c-a-b} \frac{\Gamma(c) \Gamma(a+b-c)}{\Gamma(a) \Gamma(b)}
      F(c-a, c-b, c-a-b+1; 1-z).
\end{eqnarray}
for the hypergeometric function to deduce on the asymptotic pattern of the flux in the $z\rightarrow 1$ region.
From the above analysis and from the expressions (\ref{flux1})  and  (\ref{lintransf})  it is evident that the leading  term in the expansion of the flux around $z=1$ is of the order $O({(1-z)}^{2\beta - 1/2 }),$ that is 
$O({(1-z)}^{-lm -1 })$. As far as their divergent structures are concerned, the remaining divergent terms in the flux  range from 
  $O({(1-z)}^{-lm -1 })$ to $O({(1-z)}^{-1/2 })$  and they all have the same prefactor $~ \frac{\Gamma(c)\Gamma(c -a -b)}{\Gamma(c - a)\Gamma(c - b)} ~$ which multiplies them. This can be readily seen by  combining the relations (\ref{psi1sol1}) and (\ref{lintransf}) and by inserting them into  (\ref{flux1}) and then by carefully writing out all singular terms in the expansion. All those terms in the expansion that don't have this same prefactor will anyway   vanish in the limit $~ z \rightarrow 1.$
Therefore, the vanishing of the flux in the asymptotic region imposes the following requirement,
\begin{equation}
\frac{\Gamma(c)\Gamma(c -a -b)}{\Gamma(c - a)\Gamma(c - b)} = 0.
\end{equation}
 Complying with this condition requires either $~ c-a=-n ~$ or $~ c-b=-n, ~$ with $~ n=0,1,2,...$

 Hence, in order to get the QNM's we have analyzed the behavior
of the flux  (\ref{flux1}) at  far infinity and subsequently  have  constrained it to
vanish there. As a result, there emerge two conditions
\begin{equation}\label{cond}
k_+ +k_-= -\frac{2i}{l}\left(n + \frac{lm}{2}  + \frac{1}{4} \right) \quad \text{or} \quad k_{+}-k_-= -\frac{2i}{l}\left(n   +\frac{lm}{2} + \frac{3}{4} \right).
\end{equation}
The conditions \eqref{cond} lead to the left and right QNM frequency $\omega_{L,R}$
\begin{equation}\begin{split}
&\omega_{L}=\frac{j}{l} - 2i\frac{r_+ -r_-}{l}\left(n  +\frac{lm}{2} + \frac{1}{4}\right),  \\
&\omega_{R}=-\frac{j}{l} -2i \frac{r_+ +r_-}{l}\left(n  +\frac{lm}{2} + \frac{3}{4} \right). \label{cond1}
\end{split}\end{equation}

The NC contribution is hidden in $r_\pm$ via \eqref{rplus} and \eqref{J}. Expanding up to first order in the deformation parameter $a$, we obtain
\begin{equation}
r_{+}+r_-=l\sqrt{M}+\frac{J(a)}{2\sqrt{M}} -\frac{1}{8}\frac{J^2(a)}{l M^{3/2}} + O(a^{3/2}),\quad  r_{+}-r_-=l\sqrt{M}-\frac{J(a)}{2\sqrt{M}} -\frac{1}{8}\frac{J^2(a)}{l M^{3/2}} + O(a^{3/2}).
\end{equation}
and the NC corrections to the QNM's are
\begin{equation}\begin{split}\label{NCQNM}
&\omega_{L}={\omega^{(0)}_{L}} + i \frac{1}{l\sqrt{M}} \left( J(a) + \frac{1}{4} \frac{J^2(a)}{lM} \right)  \left(n+\frac{1}{4} +  \frac{lm}{2} \right) + O(a^{3/2}),  \\
&\omega_{R}= {\omega^{(0)}_{R}} - i \frac{1}{l\sqrt{M}} \left( J(a) - \frac{1}{4} \frac{J^2(a)}{lM} \right) \left(n+\frac{3}{4} +  \frac{lm}{2} \right) + O(a^{3/2}).
\end{split}\end{equation}
where $J(a)$ is given by (\ref{J}) and $ {\omega^{(0)}_{L,R}} $ are the undeformed QNM's defined by
\begin{equation}\begin{split} \label{commuttivefreqs}
& {\omega^{(0)}_{L}}=\frac{j}{l} - 2i  \sqrt{M} \left(n+\frac{1}{4}+ \frac{lm}{2}\right),  \\
& {\omega^{(0)}_{R}}=-\frac{j}{l} -2i \sqrt{M} \left(n+\frac{3}{4}+ \frac{lm}{2}\right).
\end{split}\end{equation}
We see that  NC corrections change only
the imaginary  part of the QNM's frequencies. This is obvious when the QNM's frequencies are separated explicitly into its imaginary and real part,
\begin{equation}\begin{split}  \label{fermioqnmfreqs}
& \omega_{L}=\frac{j}{l} 
- 2i  \sqrt{M} \left( 1- \frac{J(a)}{2lM} - \frac{1}{8} \frac{J^2(a)}{l^2 M^2} \right)  \left(n+\frac{1}{4} + \frac{lm}{2} \right)
    + O(a^{3/2}),  \\
& \omega_{R}=-\frac{j}{l}  
- 2i  \sqrt{M} \left( 1+ \frac{J(a)}{2lM} - \frac{1}{8} \frac{J^2(a)}{l^2 M^2} \right)  \left(n+\frac{3}{4} + \frac{lm}{2} \right)
    + O(a^{3/2}).
\end{split}\end{equation}
These QNM's are related to the dual CFT and we will discuss more on that issue in the next section.

\section{QNM and holography} 

Holography \cite{thooft1,susskind} and AdS/CFT duality \cite{malda,Aharony} have emerged as some of the most fundamental developments in string theory and certain quantum theories of gravity, which are relevant at the Planck scale. Since NC theories are also relevant at the Planck scale \cite{dop1,dop2}, it is natural to investigate holography within the NC framework \cite{manin}. QNM's play an important role in the study of holography. In particular, the connection between QNM frequencies and the AdS/CFT duality have been studied extensively in the literature \cite{rama,horo,7,ds2}. It was shown in \cite{7,ds2} that for the BTZ black hole, the poles of the retarded Green's function in the boundary CFT are in exact correspondence with the QNM frequencies in the bulk, which provides a strong evidence for holography in the BTZ space-time. Further evidence for holography in the case of the BTZ comes from Sullivan's theorem, which says that for a certain class of manifolds, there is a 1-1 correspondence of the hyperbolic structure as encoded in the metric and the conformal structure of the boundary \cite{sullivan}. It has been shown that the Sullivan's theorem is applicable for the BTZ black hole \cite{gfin1,gfin2,gfin3}, which provides an exact kinematical statement of holography for the BTZ. Furthermore, using certain monodromy conditions which can be derived using the Sullivan's theorem, it is possible to calculate the so called nonquasinormal frequencies for the BTZ black hole, which have a form that is identical to the usual QNM frequencies for the BTZ \cite{dannynqnm,ksgnqnm}. It is thus fair to say that there is a strong connection between the AdS/CFT duality and the BTZ black hole.

As discussed in Section II, probing a spinless BTZ black hole with a NC scalar field is equivalent to probing a spinning BTZ black hole with a commutative scalar field \cite{2}. The spin of the BTZ is proportional to the NC parameter up to the first order and can be interpreted as a back reaction of the NC probe to the commutative geometry. The spinning black hole thus obtained contains information about the NC physics. Moreover, being a classical BTZ black hole, it satisfies all the conditions of holography as stated above. In particular, the results of \cite{7,ds2} relating the QNM frequencies to the poles of the retarded Green's function in the dual boundary CFT are exactly valid. The QNM frequencies for this BTZ black hole depend on the NC parameters through the spin. Thus on the gravity side, the effect of the NC physics explicitly shows up in the QNM frequencies. Using the well established AdS/CFT correspondence for the BTZ black hole \cite{7,ds2}, we can now argue that the poles of the retarded Green's function in the dual CFT would have the exact same form as the QNM frequencies and thus the dual CFT would also carry information about the NC physics. From Eqn. 
(\ref{cond}) we find that the left and right conformal weights of the dual CFT are given by $h_L = \frac{lm}{2} + \frac{1}{4}$ and $h_R = \frac{lm}{2} + \frac{3}{4}$ respectively. The poles of the retarded Green's function however pick up a NC contribution given by $2i\sqrt{M} \left ( \frac{J(a)}{2lM} + \frac{J^2(a)}{8l^2M^2} \right )$ and $-2i\sqrt{M} \left ( \frac{J(a)}{2lM} - \frac{J^2(a)}{8l^2M^2} \right ),$ respectively, where $J(a)$ is defined in (\ref{J}). These corrections to the poles of the retarded Green's function can be interpreted as arising from a NC correction to the dual CFT, up to the first order in the deformation parameter. Our work thus provides a glimpse as to what could be the nature of a NC dual CFT for the BTZ black hole. 

As discussed above, another manifestation of the holography for the BTZ black hole appears through its connection with Sullivan's theorem \cite{gfin1,gfin2,gfin3}. It was shown in \cite{dannynqnm} that instead of using the boundary conditions at infinity, certain monodromy conditions could be imposed on the solutions of a massless KG equation in the background of a BTZ black hole to give exactly the same QNM frequencies. Since this process did not involve the usual boundary condition at infinity, the resulting solutions were called non-QNM's.  We emphasize that the monodromy conditions in \cite{dannynqnm} were so chosen that the QNM and non-QNM frequencies matched exactly. It was subsequently shown in \cite{ksgnqnm} that the monodromy conditions used in \cite{dannynqnm} follow from the application of Sullivan's theorem to the BTZ black hole. Thus the idea of holography as encoded in Sullivan's theorem and applied to the BTZ space-time gives rise to the non-QNM's. It was shown in \cite{dannynqnm} that for a massless scalar field, the QNM and non-QNM for the BTZ black hole are identical. Below we show that the same holds for a massless NC fermionic field in the background of the BTZ  black hole.

We start with the  full radial components of the solution which, according to (\ref{anzatz}), read as
\begin{equation}\begin{split} \label{non-qnm1}
& \psi_1 (\rho) = \frac{1}{2} \bigg( P(\rho) + Q(\rho) \bigg) = \frac{1}{2} \frac{{(1-z)}^{1/4}}{z^{1/4}} \bigg[ \left( \sqrt{1+\sqrt{z}} + \sqrt{1-\sqrt{z}} \right) \psi'_1 + \left( \sqrt{1+\sqrt{z}} - \sqrt{1-\sqrt{z}} \right) \psi'_2 \bigg]\\
& \psi_2 (\rho) = \frac{1}{2} \bigg( P(\rho) - Q(\rho) \bigg) = \frac{1}{2} \frac{{(1-z)}^{1/4}}{z^{1/4}} \bigg[ \left( \sqrt{1+\sqrt{z}} - \sqrt{1-\sqrt{z}} \right) \psi'_1 + \left( \sqrt{1+\sqrt{z}} + \sqrt{1-\sqrt{z}} \right) \psi'_2 \bigg],
\end{split}\end{equation}
with $\psi'_1, \psi'_2$ given in (\ref{psi1sol1}). The components (\ref{psi1sol1}) themselves represent solutions around the outer horizon at $r=r_+~ (z=0)$. In what follows it is sufficient to focus only on the first component. 

Near the outer horizon, the radial part of the ingoing solution behaves as $~ z^{\alpha - 1/4},~$ leading to the monodromy 
\begin{equation} \label{non-qnm2}
 {\mathcal{M}}(r_+) = \exp \bigg[ \frac{\pi l}{r_+^2 - r_-^2} \left( \omega r_+ - \frac{jr_-}{l} \right) -i \frac{\pi}{2}  \bigg]. 
\end{equation}
The latter expression gives the change under a $2\pi$ rotation in the complex $r$ plane around the singular point at $r=r_+.$ In order to find the solution around the inner horizon, it is necessary to analytically continue the solutions (\ref{psi1sol1}) to $r=r_- ~(z=\infty).$ This is achieved by means of the linear transformation
\begin{eqnarray}  \label{lintransf1}
  F(a,b,c;z)  & = & \frac{\Gamma(c) \Gamma(b-a)}{\Gamma(b) \Gamma(c-a)} {(-z)}^{-a} F(a,1-c+a,1-b+a; \frac{1}{z})  \nonumber \\
     &  +  &  \frac{\Gamma(c) \Gamma(a-b)}{\Gamma(a) \Gamma(c-b)} {(-z)}^{-b}
      F(b, 1-c+b, 1-a+b; \frac{1}{z}).
\end{eqnarray}
The relation (\ref{lintransf1}) may be used to infer the behaviors of the two linearly independent solutions around $z=\infty,$ which turn out
to behave as $~z^{\alpha - a +1/2} {(1-z)}^{\beta}$ and $~z^{\alpha -b +1/2} {(1-z)}^{\beta}$.
Continuing in the same fashion as before, the respective monodromies are shown to be
\begin{equation}\begin{split} \label{non-qnm3}
&  {\mathcal{M}}^{+}(r_-) = \exp \bigg[  - \frac{\pi l}{r_+^2 - r_-^2} \left( \omega r_- - \frac{jr_+}{l} \right)   \bigg], \\
& {\mathcal{M}}^{-}(r_-) = \exp \bigg[   \frac{\pi l}{r_+^2 - r_-^2} \left( \omega r_- - \frac{jr_+}{l} \right) - i\pi   \bigg],
\end{split}\end{equation}
where we have set the fermion mass $m=0$.
Imposing the conditions $~{\mathcal{M}}(r_+) {\mathcal{M}}^{+}(r_-) =1~$ or $~{\mathcal{M}}(r_+) {\mathcal{M}}^{-}(r_-) =1~$, which were used in \cite{dannynqnm} and which follow from Sullivan's theorem \cite{ksgnqnm}, we get the non-QNM frequencies
\begin{equation}\begin{split} \label{non-qnm4}
&\omega=-\frac{j}{l} -2i \frac{r_+ +r_-}{l}\left(n  -\frac{1}{4}  \right), \\
&\omega=\frac{j}{l} - 2i\frac{r_+ -r_-}{l}\left(n   - \frac{3}{4}\right),
\end{split}\end{equation}
 where $n \in \mathbb{N}$. Due to $n$ being integer, it is possible in the above relations to redefine $n-1 \rightarrow n,$ so that 
a subsequent comparison with (\ref{cond1}) clearly shows that these two sets of frequencies are  the same (for $m=0$). 
We have thus shown that in the case of NC fermions probing BTZ geometry, the non-QNMs that arise from holography as applied to the BTZ geometry coincide exactly with the QNM's. This indicates that the holographic ideas continue to play an important role even in the presence of NC fields.

%%%%%%%%%%%%%%%%%%%%

\section{Quantum tunneling of Dirac particles in the presence of noncommutativity and Hawking temperature}

In order to extract the relevant information from the Eq.(\ref{diracnovo2}), we apply the WKB approximation and require the two component spinor $\Psi$ to take  the following form,
\begin{equation} \label{wkb}
  \Psi = \left(\begin{matrix} \psi_1  \\ \psi_2  \\   \end{matrix}\right) =
        \left(\begin{matrix} A(x^+, \rho, x^-)  \\  B(x^+, \rho, x^-)  \\   \end{matrix}\right)
       \exp \bigg[  \frac{i}{\hbar} S(x^+, \rho, x^-) \bigg],
\end{equation}
where the amplitudes  $~A ~$ and $~B~$ are considered to be the slowly varying functions of the coordinates $~ x^+, \rho, x^-.$ This means that  derivatives of the amplitudes $~A ~$ and $~B~$  
  are much smaller than the amplitudes itself. Besides that, after plugging  solution (\ref{wkb}) into Eq.(\ref{diracnovo2}), the terms with $\partial_{\mu } A, \partial_{\mu} B, ~$ will be of the order $ O(\hbar)  $ and thus can
  be neglected in the WKB approximation, leading to the set of equations
\begin{equation}\begin{split} \label{wkb1}
& A  ~\bigg(  \frac{l}{\cosh \rho} \frac{\partial S}{\partial x^-} - i \hbar ml \bigg)
   + B ~\bigg(  \frac{\partial S}{\partial \rho} + \frac{l}{\sinh \rho}  \frac{\partial S}{\partial x^+} \bigg) = 0,  \\
&  A ~\bigg(  \frac{\partial S}{\partial \rho} - \frac{l}{\sinh \rho}  \frac{\partial S}{\partial x^+} \bigg)
   + B ~\bigg( - i\hbar ml -  \frac{l}{\cosh \rho} \frac{\partial S}{\partial x^-}   \bigg) = 0.
\end{split}\end{equation}
This is a homogeneous system of two linear equations with two unknowns. The existence of nontrivial solutions requires
\begin{equation} \label{wkb2}
 -{\hbar }^2 m^2 l^2 - \frac{l^2}{\cosh^2 \rho} {\bigg( \frac{\partial S}{\partial x^-}  \bigg)}^2
    - {\bigg( \frac{\partial S}{\partial \rho}  \bigg)}^2 + \frac{l^2}{\sinh^2 \rho} {\bigg( \frac{\partial S}{\partial x^+}  \bigg)}^2 = 0.
\end{equation}
Due to the presence of two Killing vectors $~\partial_t \;$ and $ \; \partial_{\phi} ~$ in the BTZ spacetime, the Hamilton-Jacobi phase $~ S(x^+, \rho, x^-) = S(t, r, \phi ) ~$ can be separated as
\begin{equation} \label{wkb3}
   S(x^+, \rho, x^-) = - \frac{\omega }{l} t + \frac{j}{l} \phi + R(r) + K .
\end{equation}
Since according to relations (\ref{transformations}) and (\ref{kpluskminus}), $~ \frac{\omega }{l} t - \frac{j}{l} \phi     
      = k_+  x^+ + k_- x^-,~$ the above separation can be written as
\begin{equation} \label{wkb4}
   S(x^+, \rho, x^-) = - (  k_+  x^+ + k_- x^-) + R(\rho) + K,
\end{equation}
allowing us to rephrase Eq.(\ref{wkb2}) as
\begin{equation} \label{wkb5}
   {\bigg( \frac{\partial R}{\partial \rho}  \bigg)}^2 = 
 -{\hbar }^2 m^2 l^2 - \frac{l^2}{\cosh^2 \rho} {k_-}^2
     + \frac{l^2}{\sinh^2 \rho}  {k_+}^2.
\end{equation}

From the defining relation for the variable $~\rho, ~$  Eq.(\ref{transformations}), it is clear that its range is $~ 0 \le \rho < \infty, ~$ which corresponds to the range $~ r_+ \le r  < \infty $ in the variable $r.$ The variable $\rho$ thus cannot account for and is not able to cover the region  between the inner and the outer horizon, $~ r_- \le r  < r_+. $
This implies that the quadrature of Eq.(\ref{wkb5}), 
\begin{equation} \label{wkb6}
   R(\rho) = 
  \pm \int_0^{\rho} \sqrt{-{\hbar }^2 m^2 l^2 - \frac{l^2}{\cosh^2 \rho'} {k_-}^2
     + \frac{l^2}{\sinh^2 \rho'}  {k_+}^2}  ~d \rho',
\end{equation}
has a singularity placed at the integral's lower bound. In the context of calculating the quantum tunneling probability amplitude, such kind of singularity will lead to a result for the probability amplitude that is slightly different with respect to the amplitude obtained when one pushes the analysis exclusively in the standard coordinates $t, r, \phi.$ To see this, 
let us first note that the  probability amplitude $\Gamma$ for a quantum tunneling of fermions from inside to outside of the outer horizon is determined \cite{Mitra:2006qa},\cite{Li:2008ws} by the imaginary part, $~Im ~R ~$ of the function $R.$ 
\begin{equation} \label{wkb7}
   \Gamma = \frac{P_{out}}{P_{in}} =  \exp \bigg[  -\frac{4}{\hbar} ~Im ~R_+ \bigg],
\end{equation}
with $~ R_{\pm} ~$ referring to the upper, i.e.  lower sign in Eq.(\ref{wkb6}).
For the purpose of evaluating the integral in (\ref{wkb6}), we switch the integration variable in (\ref{wkb6})  from  $~\rho$ to $~r,$
$$
 d\rho = \frac{r dr}{r_+^2 - r_-^2} \frac{1}{\sinh \rho \cosh \rho}.
$$
leading to
\begin{equation} \label{wkb8}
   R(r) = 
  \pm \int_{r_+}^{r}  \frac{\sqrt{-{\hbar }^2 m^2 l^2 (r'^2 - r_+^2)(r'^2 - r_-^2) + l^2 (r_+^2 - r_-^2) 
    \bigg( k_+^2 (r'^2 - r_-^2) - k_-^2 (r'^2 - r_+^2)  \bigg)}}{(r'^2 - r_+^2)(r'^2 - r_-^2)}  ~r'd r'.
\end{equation}
As can be seen from figure 1, the imaginary part in the above integral comes from the contour integration over  a semicircle of radius $~ \epsilon ~$ that encircles the pole at $~ r= r_+,$
\begin{equation}\begin{split} \label{wkb9}
&  R(r) = \pm \bigg[   \int_{\gamma (\epsilon)}  \frac{\sqrt{-{\hbar }^2 m^2 l^2 (r'^2 - r_+^2)(r'^2 - r_-^2) + l^2 (r_+^2 - r_-^2) 
    \bigg( k_+^2 (r'^2 - r_-^2) - k_-^2 (r'^2 - r_+^2)  \bigg)}}{(r'^2 - r_+^2)(r'^2 - r_-^2)}  ~r'd r'  \\
&    + \lim_{\epsilon \rightarrow 0} \int_{r_+ + \epsilon}^{r}  \frac{\sqrt{-{\hbar }^2 m^2 l^2 (r'^2 - r_+^2)(r'^2 - r_-^2) + l^2 (r_+^2 - r_-^2) 
    \bigg( k_+^2 (r'^2 - r_-^2) - k_-^2 (r'^2 - r_+^2)  \bigg)}}{(r'^2 - r_+^2)(r'^2 - r_-^2)}  ~r'd r'     \bigg].
\end{split}\end{equation}
The second integral is the principal value and it is real, while the first integral gives the wanted imaginary part.
The result is
\begin{equation} \label{wkb10}
   Im ~ R_+ (\rho) = \frac{\pi}{2} \frac{\omega l r_+ - j r_-}{r_+^2 - r_-^2},
\end{equation}
leading to
\begin{equation} \label{wkb11}
   \Gamma =  \exp \bigg[ - \frac{2\pi}{\hbar} \frac{\omega l r_+ - j r_-}{r_+^2 - r_-^2} \bigg].
\end{equation}
The impact of noncommutativity scale on fermion tunneling is explicitly seen through the tunneling amplitude written in the form
\begin{equation} \label{wkb11a}
   \Gamma =  \exp \bigg[ - \frac{2\pi \omega}{\hbar \sqrt{M}} \left(  1 + \frac{3}{8} \frac{J^2}{l^2 M^2} \right)
  + \frac{\pi}{\hbar} j \frac{J}{l^2 M^{3/2}} + O(a^{3/2})   \bigg].
\end{equation}

\vspace{5.0mm}
\hspace{3cm}
 \resizebox{6.5cm}{6.5cm}{
\begin{tikzpicture}[decoration={markings,
mark=at position 2cm with {\arrow[line width=1pt]{>}},
mark=at position 8cm with {\arrow[line width=1pt]{>}},
mark=at position 18cm with {\arrow[line width=1pt]{>}},
mark=at position 29cm with {\arrow[line width=1pt]{>}},
mark=at position 32.4cm with {\arrow[line width=1pt]{>}}
}
]
% The axes
\draw[help lines,->] (-5,0) -- (5,0) coordinate (xaxis);
\draw[help lines,->] (0,-5) -- (0,5) coordinate (yaxis);
\draw[snake=zigzag] (1,0) -- (5,0);
%\draw[->] (-1,1) -- (0.8,0) ;
%\draw[->] (1,-1) -- (-0.8,0) ;

% The path
\path[draw,line width=0.8pt,postaction=decorate] (1,0.2)  -- (4,0.2)  arc (3:357:4) -- (4,-0.2) -- (1,-0.2) arc (330:30:0.4);
\filldraw (0.65,0) circle (1pt) node[below] {$$};

% The labels
\node[below] at (xaxis) {$Re ~ r'$};
\node[left] at (yaxis) {$Im ~ r'$};
%\node at (3.5,3.5) {$\Gamma_1$};
%\node at (-3.5,-3.5) {$\Gamma_2$};
%\node at (3, 0.5) {$L_1$};
%\node at (3, -0.5) {$L_2$};
\node at (-4.3, 0.4) {$-r$};
\node at (4.2, 0.4) {$r$};
\node at (0.6, 0.7) {$\gamma (\epsilon)$};
\node at (0.70,-0.7) {$r_+$};
\end{tikzpicture} }

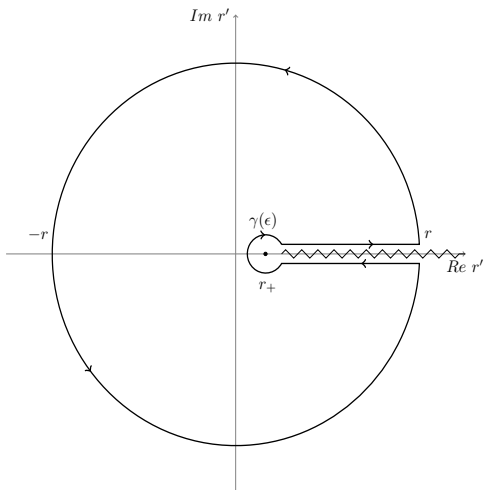
Fig.1  The radial part (\ref{wkb8}) of the classical fermion's trajectory
action $S(x^+, \rho, x^-)$ is determined by the sum of two contributions, first coming from the integration over a semicircle of radius $\epsilon$ and the other arising from the integration over the horizontal path lying above the cut. This is equivalent to taking an integral over the full circle of radius $\epsilon,$ followed by an integration along both horizontal paths, below and above the cut and then dividing by $2$. The imaginary part of (\ref{wkb8})  comes from the integration over the circle of radius $\epsilon$.

\vspace{4.5mm} 

A comparison \cite{Hartle:1976tp} with  $~ \Gamma = P_{out}/P_{in} = \exp (  -\frac{\omega}{T_H} ) ~$ gives for the Hawking temperature of BTZ black hole
\begin{equation} \label{wkb12}
   T_H  =   \frac{\hbar (r_+^2 - r_-^2)}{2\pi l r_+},
\end{equation}
that is 
\begin{equation} \label{wkb12a}
   T_H  =   \frac{\hbar  \sqrt{M}}{2\pi } \left( 1- \frac{5}{8} \frac{J^2}{l^2 M^2}  + O(a^{3/2}) \right),
\end{equation}
where the black hole spin  $J$ is given in (\ref{J}).

%For a BTZ black hole, there is an equivalent description of scalar or fermionic perturbations known as the non-quasinormal modes \cite{dannynqnm,ksgnqnm}. The calculation of the QNM's typically requires specification of the boundary conditions both at the horizon and at infinity. 
%It was shown in \cite{dannynqnm} that a particular relation between the monodromies of the fields around the black hole horizons give rise to the same frequencies and the QNM's, and a justification for such a relation based on geometric finiteness and holography was proposed in \cite{ksgnqnm}. Therefore, for certain class of space-times, the properties of the QNM's can be deduced using ideas arising from holography and string theory, relevant at the Planck scale.

%It is a well known fact that a large class of string theories contain a $AdS_3$ factor in the near-horizon region \cite{senksgmoduli} [NEED SOME MORE REFS]. The parameters of the associated BTZ black holes in the near-horizon regions of such theories  and the black hole parameters are determined in terms of the corresponding string moduli. 

\section{Final remarks}

The signatures of Planck scale physics can be encoded in the QNM's arising from the early universe. Here we have explored the effects of NC fermionic fields in the BTZ black hole background, where the Planck scale effects are modeled by the NC geometry. The corresponding QNM's have been calculated to the first order in the deformation parameter and the NC contributions have been explicitly evaluated. The analysis is based on a new duality between the NC fields and a spinless BTZ on one hand and a commutative field and a spinning BTZ on the other hand. That this duality holds was demonstrated up to the first order in the deformation parameter. While this is an approximation, the smallness of the NC deformation parameter makes it a good one and it is natural to explore the observable effects of the Planck scale physics up to this order. 

One may question why are we restricting the analysis to the first order in the deformation parameter and whether the observed duality is an artifact of that restriction. There are two primary reasons why we restricted our analysis to the first order in the deformation parameter. The physical reason is that the NC effects are supposed to be very small, primarily arising from Planck scale physics. Hence, in any possible detection of QNM's through gravitational wave laboratories, only the lowest order effects in NC parameter would be measurable, if at all. Thus from phenomenological point of view, working with lowest order is reasonable. The second reason is that during the analysis we noted that the equations of motions become very complicated even in the simple model that we are working with, namely where only the field content is taken as noncommutative. Unless we restricted ourselves to the lowest order, there would be no chance of solving the equations of motion, even in a semi-analytical fashion. At this point we wanted to make a clear and simple empirical prediction which would provide a glimpse of the possible observable consequences of the NC physics. With this goal we have restricted our analysis only to the first order.

The QNM's are known to provide an evidence of the AdS/CFT duality for the BTZ black holes. The non-QNM's were calculated in \cite{dannynqnm} using not the usual QNM boundary conditions at infinity, but rather using certain monodromy relations which were proposed in an ad hoc fashion. Subsequently it was shown in \cite{ksgnqnm} that using a particular form of holography as encoded in Sullivan's theorem \cite{gfin1}, it is possible to derive those monodromy conditions. Thus holography leads to the monodromy conditions which in turn lead to the non-QNM's. We showed that this property still survives in the NC regime. Using this duality we have also argued that our analysis provides a glimpse of the NC boundary CFT's. In the NC physics the symmetries can get deformed, mainly due to the deformed coproduct in the corresponding Hopf algebra. The role of twisted symmetries has been studied even in the context of CFT's \cite{lizzi}. In principle the dual QFT could be a twisted version of the normal CFT, but we have not yet investigated that in detail. So far we can just conjecture that the duality holds on the NC level also and that these are the corrections to the poles of retarded Green's function of NC CFT, as discussed in Section IV. The main NC effect there is to change the horizon temperatures while the conformal weights remain unchanged up to the first order in the NC deformation parameter.

Using the tunneling formalism as applied to Dirac particles in our NC framework, we have shown that the horizon temperature is lowered by the Planck scale effects. Moreover, the NC effects generally increase the tunneling amplitude, except for the lowest angular momentum channels and those with negative fermion angular momentum $j<0.$ This is due to  the emission  amplitude having two competing terms induced by Planck scale effects, which may, depending on $j$ either act constructively or may suppress each other.

BTZ black hole is a special class of space-time in 2+1 dimensions. However, it is well known that the near-horizon geometry of black holes arising from a large class of string theories contain $AdS_3$ factor, which is related to the BTZ black hole by certain discrete identifications \cite{Aharony,senksgmoduli}. Thus it is plausible that the analysis presented here is applicable to a much wider class of black holes. The parameters of such BTZ black holes in string theories are determined in terms of the string moduli. Now imagine that such a stringy black hole with zero angular momentum is being probed by a NC field. As discussed in this paper, such a probe will transform the stringy black hole with $J=0$ to a dual black hole  whose angular momentum is nonzero and is proportional to the NC parameter. The parameters of the dual BTZ black hole, which are functions of the NC parameter $a$, would then be related to the string moduli. This process would provide a link of the NC deformation parameter $a$ to the moduli space of string theories. This is consistent with the idea of NC physics appearing in string theory \cite{sw}. \\

\noindent{\bf Acknowledgment}\\
 A.S. is grateful to S.Mignemi for  discussions and  useful comments. The work of T.~J. and A.S.  has been partially supported by Croatian Science Foundation under the project (IP-2014-09-9582) and it was partially supported by the H2020 CSA Twinning project No. 692194, RBI-T-WINNING.  A.~S. would like to acknowledge the support from the European Commission and the Croatian Ministry of Science, Education and Sports through grant project financed under the Marie Curie FP7-PEOPLE-2011-COFUND, project NEWFELPRO.

\appendix

\section{Derivation of $\kappa$-deformed KG equation}

The starting point in the derivation of equation \eqref{1}, \eqref{7}  is the following action  
\begin{equation}\begin{split}\label{action1}
\hat{\mathcal{S}}&=\int \text{d}^{3}x\sqrt{-g}\ \ g^{\mu\nu}\left(\partial_{\mu}\phi\star\partial_{\nu}\phi\right)\\
&=\int \text{d}^{3}x\sqrt{-g}\ \ g^{\mu\nu}\left(\partial_{\mu}\hat{\phi}\partial_{\nu}\hat{\phi}\triangleright 1\right), \\
\end{split}\end{equation}
describing the dynamics of the NC scalar field in the  background with the classical geometry $g_{\mu \nu}$.
 The noncommutativity is introduced by replacing the usual pointwise multiplication, between the fields in the action functional, with the NC star product, i.e. $\phi(x)\phi(x)\longrightarrow\phi(x)\star\phi(x)$.

The star product is then generally given by
\begin{equation} \label{stargeneral}
(f \; \star  \; g)(x)  =   \lim_{\substack{y \rightarrow x  \\ z \rightarrow x }}
  \mu_0 \left ( e^{x^{\alpha} ( \triangle  - {\triangle}_{0}) {\partial}_{\alpha} }
    f(y) \otimes g(z) \right ),
\end{equation}
where $\mu_0 $ is the multiplication map $ \; \mu_0 (f \otimes g) = f
\cdot g \; $ and $ \; {\triangle} (\partial_\mu) \; $ is the coproduct 
for  translation generators $ \; p_\mu = i \partial_\mu \; $    and
$ {\triangle}_{0} (\partial ) = \partial \otimes 1 + 1 \otimes  \partial $
is the primitive coproduct. This structure map belongs to a
coalgebra sector of certain $\kappa$-deformation of Hopf algebra. It obviously provides
a passage where the quantum symmetry pours into the description.
We point out that
the  formula (\ref{stargeneral}) is a general one  \cite{Meljanac:2007xb,Govindarajan:2008qa,Juric:2013foa}, being valid for the
star products corresponding to any
 Lie-algebra type of deformation, of which $\kappa$-deformation is
 one particular example ( but interestingly, the
 $\theta$-deformation is not). For elucidating the origin of the formula (\ref{stargeneral}) and other
issues related to the $\kappa$-deformation, particularly those related
to the ``method of realizations'' and the correspondence between the star
product, differential operator realization, coproduct and the operator ordering prescription one may consult  \cite{Meljanac:2007xb,Govindarajan:2008qa,Juric:2013foa}.
 
When expanded up to first order in $a,$ the star product  looks as
\begin{equation} \label{starproduct}
f(x)\star g(x)=f(x)g(x)+i\beta' (\eta^{\mu \nu} x_{\mu} \frac{\partial f}{\partial x^{\nu}})(\eta^{\lambda \sigma} a_\lambda  \frac{\partial g}{\partial x^{\sigma} })+i\beta(\eta^{\mu \nu} a_\mu x_\nu)(\eta^{\lambda \sigma} \frac{\partial f}{\partial x^{\lambda}} \frac{\partial g}{\partial x^{\sigma}})+i\bar{\beta} (\eta^{\mu \nu} a_\mu  \frac{\partial f}{\partial x^{\nu}})(\eta^{\lambda \sigma} x_\lambda  \frac{\partial g}{\partial x^{\sigma}}).
\end{equation}
Here $\beta', \beta, \bar{\beta}$ are the parameters determining the
 differential operator representation of the $\kappa$-Minkowski  algebra.
% \cite{kappa}.
On the other side,  each  choice of the operator representation
corresponds \cite{Meljanac:2007xb}  to a
different choice of  coproduct (and different basis of $\kappa$-Poincar\'{e} ),  which in turn corresponds to a vacuum of the theory that  in principle should
   be fixed by experiment.
%On the other side, the choice of realization corresponds to the choice of the vacuum of the theory
%and this should be fixed by experiment, in principle.
Alongside,  $a_\mu$ is a 3-vector of deformation. In the
subsequent analysis we choose one particular orientation, $a_\mu =
(a,0,0)$, so that  the symbol $a$ from now on and in the main text is reserved for the time
component of the deformation 3-vector. This choice of  orientation
leads to the original $\kappa$-Minkowski algebra \cite{kappa1,kappa2,kappa3}.   

% In this family of realizations there is one that is particularly
% interesting. It is determined by $\beta =1$ and corresponds to the
 %phase space noncommutativity that is related to a generalized
 %uncertainty principle emerging from a study of string collisions at
 %Planckian energies 
%%\cite{gross}. 
%It is for the first time considered by Maggiore
% %\cite{genunc}.
 %It was also considered in
%% \cite{sm2, epjc, kowalski, jonke, borowiecpachol, sams}
 %where it is denoted as natural realization or classical basis.

There exists an isomorphism between the NC algebra $\cal \hat{A}$,
generated by the noncommutative coordinates 
$\hat{x}_{\mu}$ and the commutative algebra $\cal A^{\star}$,
generated by the commutative coordinates  
$x_{\mu}$, but with $\star$ as the algebra multiplication. The star product  between any two elements $f(x)$ and $g(x)$ 
in $\cal A^{\star}$ is defined as
\begin{equation}\label{star}
f(x)\star g(x)=\hat{f}(\hat{x})\hat{g}(\hat{x})\triangleright 1,
\end{equation}
where $\hat{f}(\hat{x})$ and $\hat{g}(\hat{x})$ are the elements in $\cal \hat{A}$
that are uniquely assigned to the elements $f(x)$ and $g(x),$
respectively, through the following correspondences, 
$\hat{f} (\hat{x}) \triangleright 1 = f(x), \quad \hat{g} (\hat{x})
\triangleright 1 = g(x). \; $ The element 
 $ \; 1 \; $ is the unit element in the algebra $\cal{A}$ and the action 
$\triangleright: \mathcal{H}\mapsto\mathcal{A}$ is defined by
\begin{equation}\label{djelovanje}
x_{\mu} \triangleright f(x)=x_{\mu}f(x),\quad p_{\mu}\triangleright f(x)=i\frac{\partial f}{\partial x^{\mu}}.
\end{equation} 
Here, $x_{\mu}$ and $p_{\mu}$ are the generators of the Heisenberg algebra $\mathcal{H}$  satisfying the relations,
\begin{equation}\label{H}
[x_{\mu},x_{\nu}]=[p_{\mu},p_{\nu}]=0, \quad [p_{\mu},x_{\nu}]=i\eta_{\mu\nu},
\end{equation}
where $\eta_{\mu\nu}=\text{diag}(-,+,+)$. Furthermore, the
coordinates $\hat{x}_\mu $ define $\kappa$-Minkowski algebra \cite{kappa1,kappa2,kappa3} and they
admit a differential operator representation within the enveloping algebra of
$ \mathcal{H} $  in terms of the formal power series in $x_\mu$ and
$p_\nu$. The  correspondence just described between the elements of  $\cal
\hat{A}$ and $\cal A^{\star}$ provides a ground for establishing the
isomorphism between these two structures. 

As a next step, two approximations are in order, that are motivated on the physical grounds.
The first approximation is related with the  observation
  that we are looking for the NC correction to the lowest order in the deformation parameter. The NC effects are expected to arise at the Planck scale and deformation parameter would be suppressed in powers of the Planck mass. It is therefore logical to consider the NC effects only to the lowest order.
In addition to that, we
also look at the long wavelength or low frequency limit for the solutions to
the wave equation describing the matter propagation in the background
of the black hole.
 The reason for this is that these long wavelength solutions  are associated with
 the leading contributions of the gravitational perturbations, which are inherently very weak (see  the article \cite{Kokkotas:1999bd} for a review).  There is also a considerable effort from the experimental side to detect the low frequency signals (see \cite{grwaves}).  It is therefore both logical and important to
consider the long wavelength  limit.

Following this line of arguments, after
setting in (\ref{starproduct}) $f=g=\partial\phi$, we expand the action up to  first order 
in the deformation parameter $a_{\mu}$ as
\begin{equation}\begin{split} \label{action2}
\hat{\mathcal{S}}&=\mathcal{S}_{0}+\int\text{d}^{3}x\sqrt{-g}\ \ g^{\mu\nu}\left[i\beta' x^{\sigma}\frac{\partial^{2}\phi}{\partial x^{\sigma}\partial x^{\mu}}a^{\lambda}+i\beta( \eta^{\sigma \rho} a_\sigma x_\rho)\frac{\partial^{2}\phi}{\partial x_{\lambda}\partial x^{\mu}}+i\bar{\beta} \frac{\partial^{2}\phi}{\partial x_{\alpha}\partial x^{\mu}}a_{\alpha}x^{\lambda}\right](\partial_{\lambda}\partial_{\nu}\phi), \\
\end{split}\end{equation}
where $S_0$ is the standard action functional describing the
commutative scalar field coupled to $g_{\mu \nu}$.
Since  the action in Eq.(\ref{action2}) contains the terms involving higher derivatives in the scalar 
field, that is the Lagrangian is of the general form $\mathcal{L}=\mathcal{L}(\phi,\partial \phi, \partial^{2}\phi, x),$ the
actual Euler-Lagrange equations accordingly modify,  as in the case of higher derivative theories. They read as
\begin{equation} \label{eulerlagrange}
\partial_{\mu}\frac{\delta \mathcal{L}}{\delta(\partial_{\mu}\phi)}-\partial_{\mu}\partial_{\nu}\frac{\delta \mathcal{L}}{\delta(\partial_{\mu}\partial_{\nu}\phi)}=\frac{\delta \mathcal{L}}{\delta \phi}.
\end{equation}
The equation of motion following from (\ref{eulerlagrange}) is further subject to the
long wavelength  approximation, where we keep terms in the equations of motion that are of the lowest order in derivatives. In this approximation the terms dependent on $\beta'$ 
and $\bar{\beta}$  do not contribute since they are all proportional to
  higher derivatives of the scalar field. Consequently, in the lowest order of the long 
wavelength approximation only terms depending on $\beta$
  will contribute  to that part of the equation of motion that is induced  by the noncommutativity. 
With the above  two approximations, the equation of motion reduces to
the form given in \eqref{1}, \eqref{7}. 

%%%%%%%%%%%

Salient features of the model \cite{1,2} are encoded within the radial equation  of the form 
\begin{equation}\label{eomradial}
r\left(M-\frac{r^2}{l^2}\right)\frac{\partial^2 R}{\partial r^2}+\left(M-\frac{3r^2}{l^2}\right)
\frac{\partial R}{\partial r}+\left(\frac{s^2}{r}-\omega^2\frac{r}{\frac{r^2}{l^2}-M}-a\beta\omega\frac{8r}{l^2}\frac{\frac{3r^2}{2l^2}-M}{\frac{r^2}{l^2}-M}\right)R=0,
\end{equation}
which is the radial component of the field equation for the NC scalar field and can be inferred by using the
ansatz  $\phi(r,\theta,t)=R(r)e^{-i\omega t}e^{is\theta}$ in \eqref{1}, \eqref{7}, as long as $M>>1$, and keeping terms up to first order in
the deformation parameter $a$. The parameter $s$ is the angular momentum of the scalar probe.

The  entropy of the black hole within the model considered here is obtained by using the ``brick-wall model''  \cite{gthooft}. This approach was applied to BTZ case in \cite{kim}. Following the same line of arguments as in \cite{gthooft, kim}, it is possible to get  from Eqn. \eqref{eomradial} the $r$-dependent radial wave  number in the form
\begin{equation}\label{k}
k^2 (r,s,\omega)=-\frac{s^2}{r^2\left(\frac{r^2}{l^2}-8GM\right)}+\omega^2\frac{1}{\left(\frac{r^2}{l^2}-8GM\right)^2}+a\beta\omega\frac{8}{l^2}\frac{\frac{3r^2}{2l^2}-8GM}{\left(\frac{r^2}{l^2}-8GM\right)^2},
\end{equation}
where the ansatz $R(r)=\text{e}^{i\int k(r)\text{d}r}$ and WKB approximation have been used. According to the semi-classical quantization rule, the radial wave number is quantized as
\begin{equation}
\pi n(\omega, s) =\int^{L}_{r_{+}+h} k(r,s,\omega)\text{d}r
\end{equation}
where the quantum numbers $n(\omega, s)>0$, $s$ should be fixed such that  $k(r,s,\omega)$ is real and $h$ and $L$ are ultraviolet
and infrared regulators, respectively (in what follows  we take the limit $L\rightarrow\infty$ and set $h\approx 0$ and we keep only the most divergent terms in $h$). The total 
number $\nu(\omega)$ of solutions with energy not exceeding $\omega$ is then  given by
\begin{equation}
\nu (\omega) =\sum^{s_{0}}_{-s_{0}}n(\omega, s)=\int^{s_{0}}_{-s_{0}}\text{d}s ~n(\omega, s)=\frac{1}{\pi}\int^{s_{0}}_{-s_{0}}\text{d}s\int^{L}_{r_{+}+h} k(r,s,\omega)\text{d}r.
\end{equation}
The free energy at inverse temperature $\beta_{T}$ of the black hole is 
\begin{equation}
 F= -\int_0^{\infty} \frac{\nu (\omega)  \text{d}\omega}{\text{e}^{\beta_{T}\omega}-1}=-\frac{1}{\pi}\int^{\infty}_{0}\frac{\text{d}\omega}{\text{e}^{\beta_{T}\omega}-1}\int^{L}_{r_{+}+h}\text{d}r\int^{s_{0}}_{-s_{0}}\text{d}s\ \ k(r,s,\omega).
\end{equation}
After carrying out the integrations and keeping the most divergent terms in $h$, one gets
\begin{equation}
F=-\frac{l^{\frac{5}{2}}}{(8GM)^{\frac{1}{4}}}\frac{\zeta(3)}{\beta^3_{T}}\frac{1}{\sqrt{2h}}-2a\beta\frac{(8GM)^{\frac{3}{4}}\sqrt{l}}{\sqrt{2h}}\frac{\zeta(2)}{\beta^2_{T}},
\end{equation}
which is the exact result in the sense of the  WKB method and $\zeta$ is the  Euler-Riemann zeta function.

The entropy for the NC massless scalar field  now follows from 
$S=\beta^2_{T}\frac{\partial F}{\partial \beta_{T}}$, yielding
\begin{equation}\begin{split}
S&=3\frac{l^{\frac{5}{2}}}{(8GM)^{\frac{1}{4}}}\frac{\zeta(3)}{\beta^2_{T}}\frac{1}{\sqrt{2h}}+4a\beta\frac{(8GM)^{\frac{3}{4}}\sqrt{l}}{\sqrt{2h}}\frac{\zeta(2)}{\beta_{T}}\\
&=S_{0}\left(1+\frac{4}{3}a\beta\frac{8GM}{l^2}\frac{\zeta(2)}{\zeta(3)}\beta_{T}\right).
\end{split}\end{equation}

%%%%%%%%%%%%%

\section{Another dual picture $(M^f(a), J^f=0)$}

So far we have used the entropy-equivalence in section II to obtain the dual picture where only the spin was rescaled $\left|J^f\right|\propto\sqrt{a\beta}$. But, we can demand that only the mass $M^f$ of dual setting changes. In doing so, we get 
\begin{equation}
r^{f}_{+}=l\sqrt{M^f}, \quad r^{f}_{-}=0, \quad M^f=M\left(1+a\beta \frac{\sqrt{M}}{l} \frac{16\pi}{3}\frac{\zeta{(2)}}{\zeta{(3)}}\right)+O(a^2).
\end{equation}
The surface gravity is given by
\begin{equation}
\kappa^f=\frac{r^{f}_+}{l^2}=\kappa_0\left(1+a\beta \frac{\sqrt{M}}{l} \frac{8\pi}{3}\frac{\zeta(2)}{\zeta{(3)}}\right)+O(a^2)=\kappa_{NC},
\end{equation}
with $\kappa_0 = \frac{\sqrt{M}}{l}$ being the surface gravity of the undeformed spinless BTZ black hole.
 	Moreover, the tunneling probability is
\begin{equation}
\Gamma_{NC}=\Gamma_0 \left(1+a\beta\frac{4\pi^2 l}{3}\frac{\zeta(2)}{\zeta(3)}\right)
\end{equation}
 and Hawking temperature is given by
\begin{equation}
T_{NC}=T_0 \left(1+a\beta\sqrt{M}\frac{2\pi}{3}\frac{\zeta(2)}{\zeta(3)}\right),
\end{equation}
where $T_0 = \frac{\kappa_0}{2\pi}$ is the Hawking temperature for a BTZ black hole in the absence of deformation and $\Gamma_0 = \exp \bigg[  -2 \pi \frac{\omega}{\kappa_0} \bigg] $.
The NC corrections to fermionic QNM modes are inferred from
\begin{equation}\begin{split}
&\omega_{L}=\omega_{L}^{(0)} -i \frac{16\pi}{3}  a\beta \frac{M}{l}\frac{\zeta(2)}{\zeta(3)} \left(n+\frac{1}{4}+\frac{lm}{2}\right) +O(a^2),  \\
&\omega_{R}=\omega_{R}^{(0)} -i \frac{16\pi}{3} a\beta \frac{M}{l}  \frac{\zeta(2)}{\zeta(3)} \left(n+\frac{3}{4}+\frac{lm}{2}\right) +O(a^2),
\end{split}\end{equation}
where $~\omega_{R}^{(0)},\omega_{L}^{(0)}~$ are the undeformed frequencies given by (\ref{commuttivefreqs}). We see that in this dual picture the corrections are even more suppressed.

\end{document}